\documentclass[pdflatex,sn-mathphys-num]{sn-jnl}%

\usepackage{graphicx}%
\usepackage{multirow}%
\usepackage{amsmath,amssymb,amsfonts}
\usepackage{amsthm}%
\usepackage{mathrsfs}%
\usepackage[title]{appendix}%
\usepackage{xcolor}%
\usepackage{textcomp}%
\usepackage{manyfoot}%
\usepackage{booktabs}%
\usepackage{algorithm}%
\usepackage{algorithmicx}%
\usepackage{algpseudocode}%
\usepackage{listings}%
\usepackage{hyperref}
\usepackage[english]{babel}
\usepackage{natbib}
\usepackage{amsmath}
\usepackage{graphicx}
\usepackage{booktabs}

\DeclareMathOperator{\sech}{sech}

\newcommand{\p}{\partial}
\newcommand{\rmi}{\mathrm{i}}
\newcommand{\rmm}{\mathrm{m}}

\title{Nonlinear Reshaping of Gaussian Water Wave Packets}

\author*[1]{Zibo Zheng}\email{zibo.zheng@oist.jp}
\author[2]{Takuji Waseda}
\author[3,4]{C. Martijn de Sterke}
\author[5]{Andrea Blanco-Redondo}
\author*[1,2,6,7]{Amin Chabchoub}\email{amin.chabchoub@oist.jp} 

\affil[1]{\orgdiv{Marine Physics and Engineering Unit}, \orgname{Okinawa Institute of Science and Technology}, \city{Onna-son}, \state{Okinawa 904-0495}, \country{Japan}}

\affil[2]{\orgdiv{Graduate School of Frontier Sciences}, \orgname{The University of Tokyo}, \city{Kashiwa}, \state{Chiba 277-0882}, \country{Japan}} 

\affil[3]{\orgdiv{Institute of Photonics and Optical Science (iPOS), School of Physics}, \orgname{The University of Sydney}, \city{Sydney}, \state{NSW 2006}, \country{Australia}} 

\affil[4]{\orgdiv{ARC Centre of Excellence for Optical Microcombs for Breakthrough Science (COMBS)}, \orgname{The University of Sydney}, \city{Sydney}, \state{NSW 2006}, \country{Australia}}

\affil[5]{\orgdiv{CREOL, The College of Optics \& Photonics}, \orgname{University of Central Florida}, \city{Orlando},
\state{FL 32816}, \country{USA}}

\affil[6]{\orgdiv{Department of Civil and Environmental Engineering}, \orgname{Imperial College London}, \city{London}, \state{London SW7 2AZ}, \country{UK}} 

\affil[7]{\orgdiv{Department of Infrastructure Engineering}, \orgname{The University of Melbourne}, \city{Parkville}, \state{3010 VIC}, \country{Australia}}

\begin{document}
\maketitle
\begin{abstract}

The propagation of narrowband wave packets is a classical problem with broad relevance across nonlinear wave physics. Within the framework of the nonlinear Schrödinger equation, solitons are known to propagate without changing shape, whereas Gaussian envelopes may undergo substantial reshaping depending on the balance between dispersion and nonlinearity. Here, we combine laboratory experiments with numerical simulations to systematically investigate the evolution of hydrodynamic Gaussian wave packets over a wide range of amplitudes and spectral bandwidths. We show that when the envelope amplitude and width match those of the stationary NLSE soliton solution, the wave packet propagates steadily in the wave flume. More generally, by varying the amplitude of the Gaussian envelope, the wave packet may undergo either dispersive broadening or nonlinear focusing, resembling the dynamics of Satsuma-Yajima breathers. These findings may facilitate the controlled generation and manipulation of localized wave packets in a variety of nonlinear dispersive systems, including optics, plasmas, and Bose-Einstein condensates. 

\end{abstract}

\section*{Introduction}
The nonlinear Schrödinger equation (NLSE) serves as a simple yet universal mathematical framework for describing the complex evolution of narrowband wave packets in weakly nonlinear and dispersive media \cite{zakharov1968stability,ablowitz2011nonlinear,carretero2024nonlinear}. A fundamental feature of the NLSE is its integrability and the existence of exact coherent wave packets, which may be either stationary or pulsating. Indeed, numerous experiments across different physical systems have demonstrated the robust and shape-preserving propagation of sech-type envelope solitons governed by the focusing NLSE, owing to the exact balance between dispersion and nonlinearity \cite{yuen1975nonlinear,slunyaev2013simulations,hasegawa1973transmission}. Lately, a focus has also been on exact analytical solutions of the NLSE, ranging from pulsating envelope solitons to breather solutions \cite{Chabchoub2011rogue,Chabchoub13,dudley2014instabilities}. The dynamics of envelope solitons with varying amplitudes and velocities are well understood, as the inverse scattering transform and Darboux transformation provide rigorous theoretical frameworks for predicting their evolution \cite{Zakharov70,Satsuma74,akhmediev1997nonlinear}. More generally, weakly nonlinear theory predicts that, under appropriate initial or boundary conditions, a wave packet will fission into one or more solitons over sufficiently long propagation distances, as experimentally demonstrated for rectangular and hydrodynamic multi-solitons \cite{Su82,Chabchoub13}.

Despite their theoretical elegance and the potential for considerable control over the dynamics of sech-type envelope solitons, Gaussian wave packets remain the preferred choice in practical hydrodynamic applications \cite{Kinsman65}, particularly for studies of wave-structure interaction \cite{Clauss86}, wave focusing \cite{Shemer10}, wave breaking \cite{Rodas14,Pizzo19}, wave-turbulence interaction \cite{Xuan24}, and wave-bottom interaction \cite{li2021surface}. In fact, Gaussian envelopes provide a convenient representation of ocean wave groups, including extreme wave events. However, unlike sech-type solitons, Gaussian envelopes are not stationary solutions of the NLSE and may therefore undergo significant reshaping during propagation. To address dispersive broadening, previous works have used linearly chirped packets \cite{Waseda05,dudley2007self,Agrawal13,Pizzo19} by introducing a specific phase modulation (or chirp). If this chirp has the appropriate sign, linear dispersion compresses the wave packet such that it recovers its prescribed temporal shape at a specific target location. Although this technique effectively compensates for linear dispersive broadening, it cannot eliminate shape distortions induced by strong nonlinearity.

Numerical and laboratory investigations on manipulating and controlling Gaussian-type wave envelopes have been reported, particularly in optics \cite{Agrawal13,blanco2016pure,Tam19,runge2020pure}. In hydrodynamics, previous studies have dealt with the evolution properties of Gaussian envelopes within the framework of the hydrodynamic NLSE \cite{Adcock09} and demonstrated the applicability of the modified nonlinear Schrödinger equation (MNLSE) \cite{dysthe1979note} for predicting the evolution of emerging steep wave packets \cite{Shemer10}.

In this work, we systematically investigate three distinct Gaussian envelope propagation regimes, namely {\it broadening}, {\it focusing}, and the {\it steady} case; this latter case is motivated by earlier work in nonlinear optics demonstrating the emergence of a pure-quartic soliton from Gaussian initial conditions within the framework of a generalized NLSE \cite{Tam19}. To this end, we combine deep-water  experimental wave profile measurements with numerical simulations based on the NLSE \cite{zakharov1968stability,osborne2010nonlinear} and, for strongly focusing wave groups, the MNLSE \cite{dysthe1979note,Goullet11}. Using the stationary sech-type envelope soliton of the NLSE as a reference, we show that a Gaussian wave packet with appropriately matched amplitude and width can exhibit remarkably similar propagation dynamics. Our experiments also demonstrate that Gaussian wave packets may broaden, focus, or propagate in a nearly steady manner, depending on the balance between dispersion and nonlinearity resulting from variations in the wave packet amplitude. The NLSE simulations provide excellent quantitative predictions of the envelope evolution, except in the strongly focusing regime, where higher-order effects become significant. In this regime, the MNLSE successfully captures both the asymmetric deformation of the wave packet and the associated increase in its propagation speed.


\section*{Results}
We conducted a series of experiments evoloving Gaussian wave packets at a constant water depth of $h=5$~m to ensure the applicability of the deep-water NLSE, with water-depth-independent coefficients, and report here only the eight most relevant experiments. The experiments are categorized into {\it three} regimes, with the packets' amplitudes adjusted by a factor $N$ ($0.36 \leq N \leq 1.8$; see Eq.~(\ref{eq:gauss})) while retaining the envelope soliton width: {\it dispersion-dominated} ($N=0.36$ and $0.6$), {\it quasi-stationary/solitonic} ($N=0.8$ and $1.0$), and {\it focusing/multi-soliton} ($N=1.2$, $1.4$, $1.6$, and $1.8$). Within the focusing/multi-soliton regime, the largest values of $N$, i.e., $N=1.6$ and $1.8$, correspond to extreme cases characterized by strong wave focusing, for which departures from the idealized NLSE predictions are expected due to increasingly significant higher-order effects. The relative strength of dispersion and nonlinearity can be quantified by the ratio $L_\mathrm{D}/L_\mathrm{NL}=1.12N^2$, where $L_\mathrm{D}$ and $L_\mathrm{NL}$, defined in Eq.~\eqref{eq:ld}, denote the characteristic second-order dispersion and nonlinear lengths, respectively. These length scales characterize the propagation distances over which dispersive and nonlinear effects become significant. A ratio $L_\mathrm{D}/L_\mathrm{NL}\ll1$ indicates dispersion-dominated propagation, whereas $L_\mathrm{D}/L_\mathrm{NL}\gg1$ indicates the dominance of nonlinear effects.

All parameters related to the Gaussian envelope generation for the laboratory experiments and numerical simulations, including amplitude $a$, wave frequency $\omega$, wavenumber $k$, the ratio $L_\mathrm{D}/L_\mathrm{NL}$ and dimensionless envelope amplitude control parameter $N$, are given in Table \ref{tab:experimentalParameter}. Since the smallest wavenumber is $k=2.4\,\mathrm{m}^{-1}$, the dimensionless water depth satisfies $kh\geq12$, ensuring that the deep-water condition is met, given the dispersion relation $\omega=\sqrt{gk \tanh kh}\approx\sqrt{gk}$. The values of the ratio $L_\mathrm{D}/L_\mathrm{NL}$ for the experimental cases are detailed in Table \ref{tab:experimentalParameter}, ranging from 0.15 for the highly dispersive case to 3.63 for the strongly nonlinear case. For comparison, numerical simulations are carried out using the NLSE and the MNLSE. More details on these frameworks, including the numerical scheme adopted, can be found in the \hyperref[sec:methods]{Methods} section.
\begin{table}[h]
    \centering
    \caption{Parameters used in laboratory experiments and numerical simulations.}
    \label{tab:experimentalParameter}
    \begin{tabular}{ccccc}
        \hline\hline
        wave amplitude $a\ (\rmm)$ & wave frequency $\omega\ (\rm s^{-1})$ & wavenumber $k\ (\rmm^{-1})$ & Amplitude parameter $N$ & $L_\mathrm{D}/L_\mathrm{NL}$ \\
        \hline
        0.100 & 4.85 & 2.4  & 0.36 & 0.15 \\
        0.040 & 6.26 & 4    & 0.6  & 0.40 \\
        0.012 & 8.09 & 6.67 & 0.8  & 0.72 \\
        0.012 & 8.09 & 6.67 & 1    & 1.12 \\
        0.012 & 8.09 & 6.67 & 1.2  & 1.61 \\
        0.012 & 8.09 & 6.67 & 1.4  & 2.20 \\
        0.012 & 8.09 & 6.67 & 1.6  & 2.87 \\
        0.012 & 8.09 & 6.67 & 1.8  & 3.63 \\
        \hline\hline
    \end{tabular}
    \end{table}

\subsection*{Dispersion-Dominated Regime}

\begin{figure}
\centering
\includegraphics[width=\linewidth]{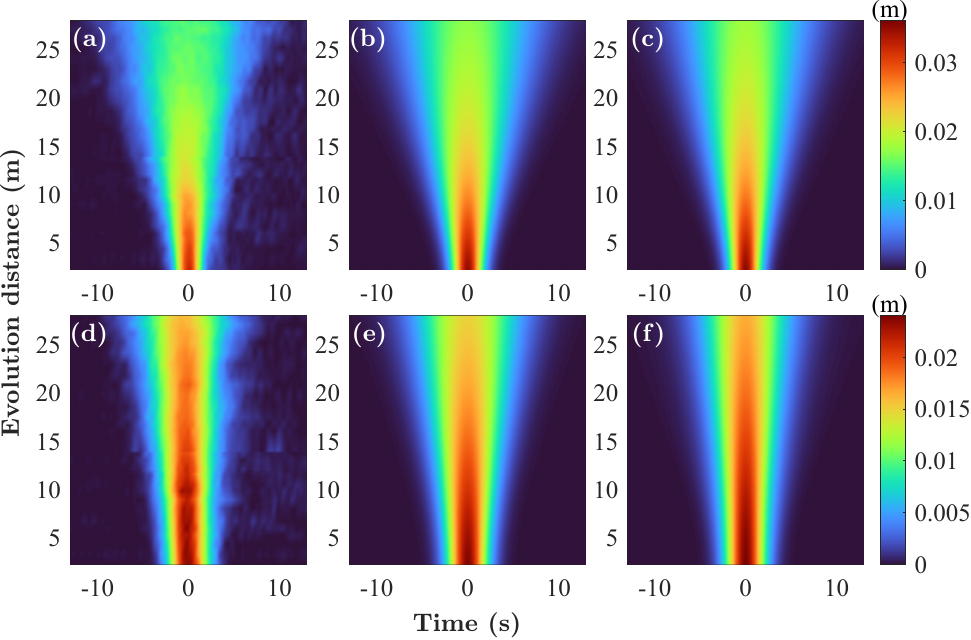}
\caption{Spatial evolution of Gaussian envelopes in the dispersion-dominated regime. (a)-(c) Evolution for the initial amplitude parameter $N=0.36$. (d)-(f) Evolution for $N=0.6$. Panels (a) and (d) show the experimental results, (b), (e) show numerical results using the LSE, whereas (c), (f) show numerical results from the NLSE. All results are presented in a reference frame moving with the group velocity $c_g$.} 
\label{fig:broaden}
\end{figure}

The comparison between the experimental and numerical evolution of the temporal pulse shape is given in Fig. \ref{fig:broaden} for $N=0.36$ and $0.6$. The corresponding evolution of the envelope amplitude is given in Fig. \ref{fig:ampvsx}(a). Excellent agreement with the conservative prediction, as described by Eq. (\ref{eq:nls}), is observed for both cases. The simulations with the linear Schr\"{o}dinger equation (LSE) also predict the evolution of these both latter cases well, considering the ratio of the dispersion length to the nonlinear length, quantifying the relative importance of nonlinearity over dispersion, being 0.15 and 0.40 for $N=0.36$ and $N=0.6$, respectively, see Table \ref{tab:experimentalParameter}. The weakly nonlinear case corresponding to $N=0.36$, shown in Figs.~\ref{fig:broaden}(a)-(c), propagates over a distance of $3.45L_\mathrm{D}$ and exhibits substantial reshaping of the wave envelope, as expected in the dispersion-dominated regime.
The envelope amplitude decreases monotonically by approximately 50\% at the last wave gauge, while its width increases by approximately a factor of four, thus, the wave energy is approximately conserved. 

A similar behavior can be observed when we steepen the wave packet and tune the amplitude parameter to $N=0.6$. The respective hydrodynamic evolution is shown in Figs.~\ref{fig:broaden}(d)-(f). Upon propagating downstream the water channel over $2.55L_\mathrm{D}$, the wave packet also decreases in amplitude while becoming broader. However, this deformation is less severe than in the case $N=0.36$. This is due to the increased wave packet amplitude imposed by the boundary conditions, which causes nonlinear effects to increasingly counterbalance dispersion, thereby delaying the broadening of the packet. Consequently, the LSE simulation predicts a faster decay of the envelope amplitude than observed in the experimental measurements and the corresponding NLSE predictions. Here, the propagation distance with respect to the characteristic dispersion length $L_\mathrm{D}$ for $N=0.6$ is smaller than for $N=0.36$, which also makes the decay of the wave packet less pronounced over the same physical distance.

\begin{figure}
\centering
\includegraphics[width=\linewidth]{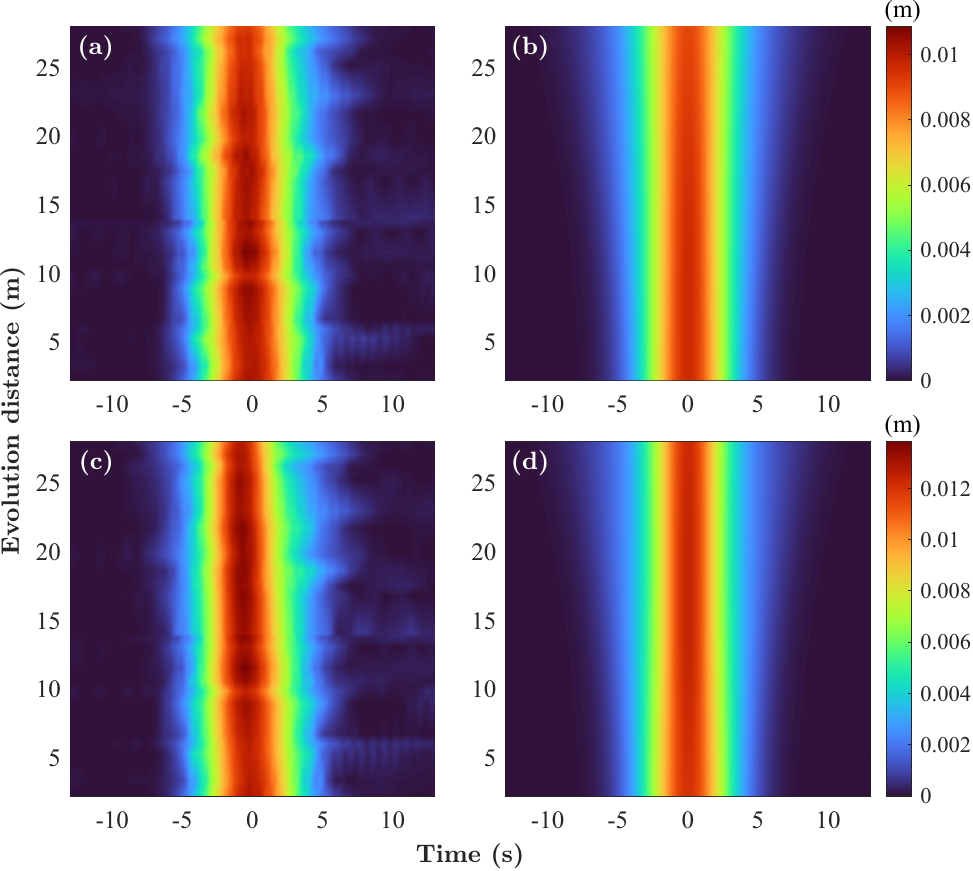}
\caption{Spatial evolution of Gaussian envelopes in the quasi-stationary and solitonic regime. (a), (b) Evolution for the initial amplitude parameter $N=0.8$. (c), (d) Evolution for $N=1.0$. (a) and (c) show the experimental results, whereas (b) and (d) refer to the respective numerical NLSE results. All results are presented in a reference frame moving with the group velocity $c_g$.}
\label{fig:quasistable}
\end{figure}

\subsection*{Quasi-Stationary and Solitonic Regime}
As the initial amplitude parameter $N$ approaches 1, our experiments demonstrate that the wave packets evolve into a quasi-stationary state, as shown in Figs.  \ref{fig:quasistable}(a) and (b). The wave packets maintain relatively robust profiles, with only slight variations in amplitude and width observed over a propagation distance of $1.07L_\mathrm{D}$.

An excellent agreement in the amplitude evolution can also be observed for $N=1$. Theoretically, this initial condition converges to an exact sech-type envelope solution of the NLSE. The discrepancy observed between the experiments and numerical simulations arises from higher-order effects, including self-steepening and wave-induced mean flow drift, which leads to an amplitude-dependent group velocity, both of which are key characteristics of steep wave group dynamics that are not captured by the third-order approximation, $\mathcal{O}(\varepsilon^3)$, of the water wave problem underlying the NLSE framework \cite{dysthe1979note,Chabchoub13,shemer2013peregrine,Pizzo16}. These effects become more pronounced as the wave group undergoes focusing, which we will discuss next.

\begin{figure}
\centering
\includegraphics[width=\linewidth]{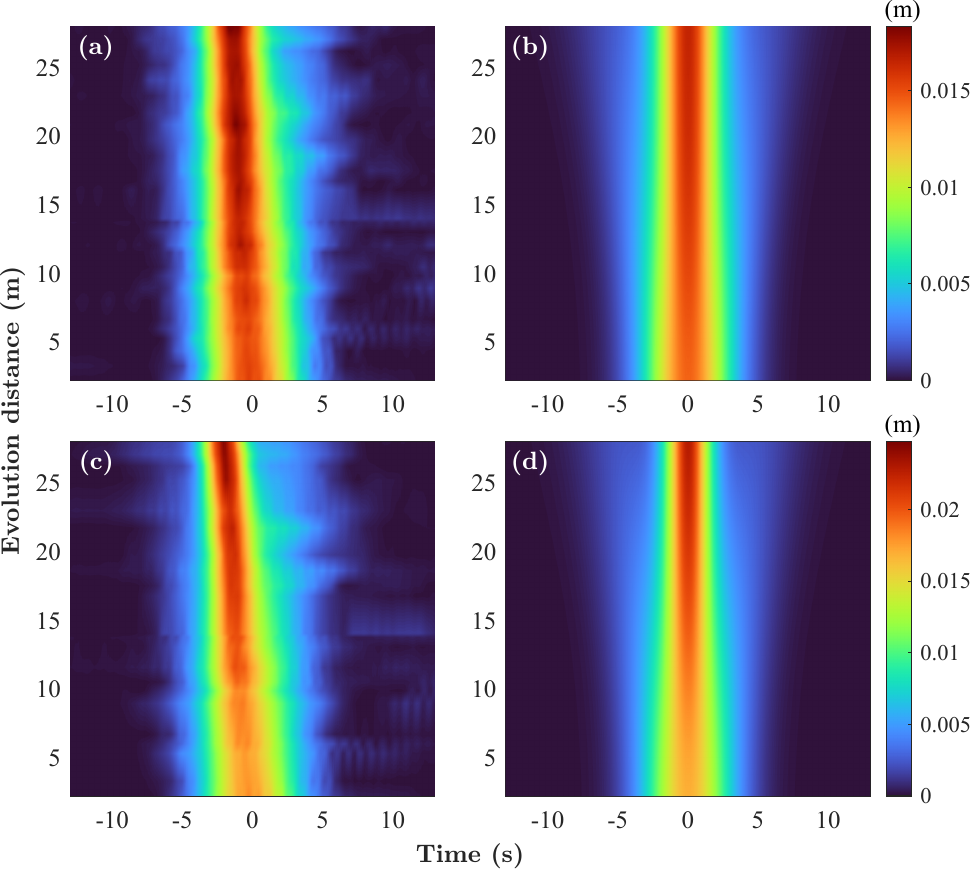}
\caption{Spatial evolution of Gaussian envelopes in the focusing and multi-soliton regime. (a) and (b) Evolution for the initial amplitude parameter $N=1.2$. (c) and (d) Evolution for $N=1.4$. Panels (a) and (c) show the experimental results, while panels (b) and (d) refer to the respective numerical NLSE results. All results are presented in a reference frame moving with the group velocity $c_
g$.}
\label{fig:steepen}
\end{figure}



\subsection*{Focusing and Multi-Soliton Regime}

Finally, we consider the cases $N=1.2$ and $1.4$, for which nonlinear effects dominate over dispersion. The resulting self-focusing of the wave packets is clearly observed during propagation, as shown in Fig.~\ref{fig:steepen}. In both cases, the wave packet profiles exhibit a progressive increase in amplitude accompanied by a decrease in width.

For $N>1$, the imposed boundary conditions initiate an unsteady wave packet for which nonlinear effects increasingly dominate over dispersion, resulting in self-focusing and substantial envelope reshaping. As the wave packet propagates, its peak amplitude increases, the envelope narrows, and its peak travels faster than the linear group velocity $c_g$. For $N$ slightly above unity, these features are captured reasonably well by the NLSE simulations, although departures from quantitative agreement with the experiments begin to emerge.

\begin{figure}
\centering
\includegraphics[width=\linewidth]{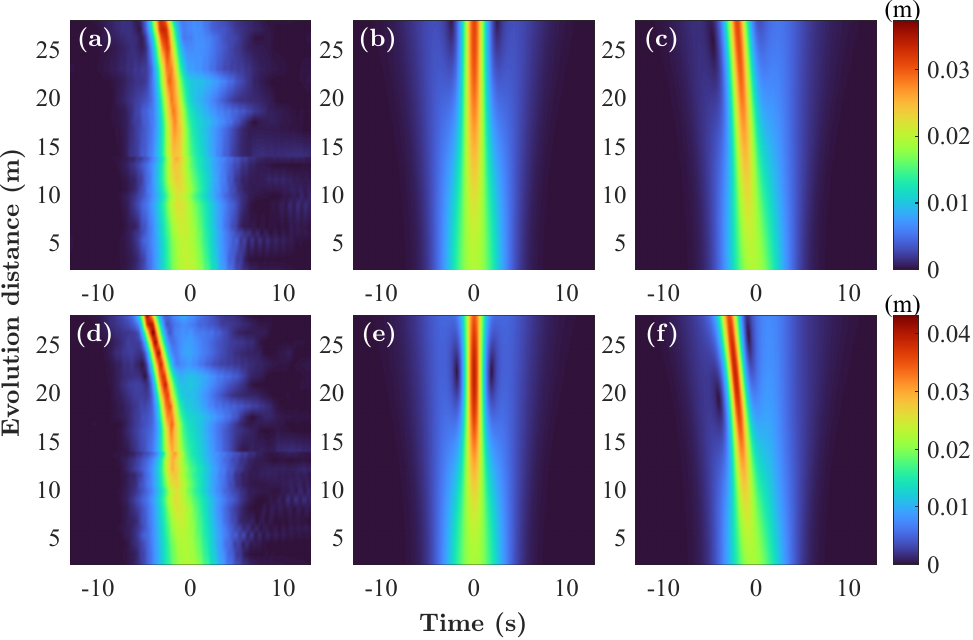}
\caption{Spatial evolution of Gaussian envelopes in the focusing and multi-soliton regime. (a)-(c) Evolution for the initial amplitude parameter $N=1.6$. (d)-(f) Evolution for $N=1.8$. Panels (a, d) show the experimental results, while panels (b), (e) and (c), (f) show the numerical results with the NLSE and the MNLSE, respectively. All results are presented in a reference frame moving with the group velocity $c_g$.}
\label{fig:split}
\end{figure}

As $N$ is increased further and nonlinearity becomes increasingly dominant, the same focusing process becomes considerably more pronounced, as shown in Fig.~\ref{fig:split}. Compared with the moderately focusing cases in Fig.~\ref{fig:steepen}, the envelope focuses more rapidly, reaching a larger peak amplitude and a narrower width.
At the same time, the increased spectral bandwidth and higher-order nonlinear effects become non-negligible, leading to dynamics that are no longer adequately described by the NLSE \cite{Chabchoub13}. In particular, the experiments exhibit an asymmetric envelope profile and a substantial deviation of the envelope-peak velocity from the linear group velocity $c_g$. The latter effects are associated with higher-order nonlinear dispersion and wave-induced mean flow. These contributions extend beyond the $\mathcal{O}(\varepsilon^3)$ approximation of the NLSE. We therefore resort to the MNLSE, Eq.~(\ref{eq:mnlse}), which incorporates these higher-order effects at $\mathcal{O}(\varepsilon^4)$ \cite{dysthe1979note}. As shown in Figs.~\ref{fig:split}(c) and (f), the MNLSE provides a substantially improved description of the strongly nonlinear wave evolution, accurately capturing the asymmetric envelope deformation, the increased propagation speed, the onset of soliton fission, and the evolution of the packet width. 

For the most extreme case $N=1.8$, which approaches the dynamics of a second-order Satsuma-Yajima soliton, the amplitude is enhanced by approximately 80\% at the last gauge and as shown in Fig. \ref{fig:ampvsx} (d). Amplitude decay can also be observed in this case from the simulation, which agrees with the oscillating behavior shown in Fig. 2 of \citep{Satsuma74}. Theoretically, the curves in Fig. \ref{fig:ampvsx}(c) should also exhibit such behavior. As shown in \citep{Satsuma74}, larger $N$ leads to a smaller period beating, which cannot be observed for $N<1.5$ due to the limited length of the water tank. Also in this case, the complete focusing-defocusing cycle of the wave envelope cannot be observed in the wave tank, due to the choice of carrier wave parameters and limited wave tank length.

\section*{Discussion}
Unlike sech-type envelope solitons, Gaussian wave groups are not exact stationary solutions of the NLSE and therefore do not inherently preserve a balance between dispersion and nonlinearity. Nevertheless, our experiments demonstrate that their evolution can be classified into three distinct dynamical regimes, all of which can be quantitatively described by the NLSE or, when higher-order effects become significant, by the MNLSE.

The most remarkable experimental finding is the observation that a Gaussian envelope can remain nearly stationary over a propagation distance of $1.07L_\mathrm{D}$, provided that its spectral width and amplitude are approximately equal to those of the corresponding NLSE envelope soliton. This result highlights the robustness of nonlinear self-organization mechanisms and suggests that Gaussian initial conditions may naturally evolve toward soliton-like attractors under suitable conditions.

If this balance condition is not satisfied, that is, when the Gaussian envelope is initialized with the same bandwidth but with an amplitude different from that of the corresponding envelope soliton, two additional evolution scenarios may emerge. In the present study, only the amplitude is varied while the spectral bandwidth is kept fixed. Equivalently, one could fix the amplitude and vary the bandwidth. For amplitudes smaller than the sech-type soliton reference, the Gaussian envelope experiences gradual attenuation and an early stage of reshaping during propagation for $0.5<N<1.0$. In contrast, larger amplitudes lead to pronounced nonlinear focusing dynamics resembling multi-soliton interactions and strongly localized wave amplification. 

\begin{figure}
\centering
\includegraphics[width=\linewidth]{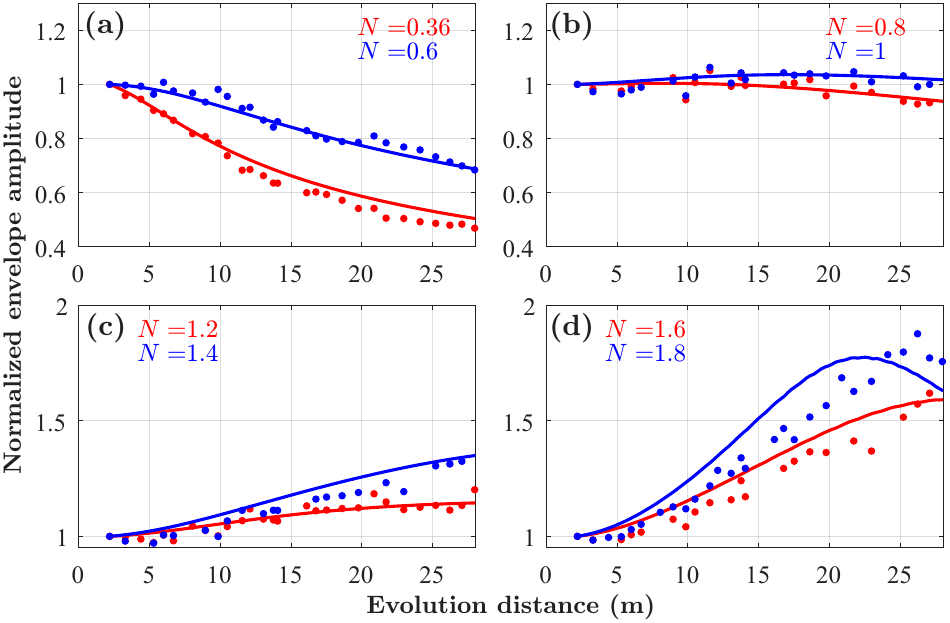}
\caption{Evolution of normalized envelope amplitude along the water channel. The panels show the evolution for (a) the dispersion-dominated, (b) the quasi-stationary and solitonic, and (c),(d) the focusing and multi-soliton regimes. Solid lines represent the numerical simulations with the NLSE in (a)-(c) and the MNLSE in (d), while dotted lines represent the experimental data.}
\label{fig:ampvsx}
\end{figure} 

Qualitatively, the initial reshaping of the wave profiles of the three regimes shows behaviors similar to the early-stage evolution in Figs. 1-3 of \cite{Su82}. Although rectangular envelopes were adopted in that work, the sharp gradients quickly smooth out, naturally initializing into well-controlled pulse shapes. Regarding the long-time evolution, we anticipate that over sufficiently long propagation distances, the Gaussian envelope will either disperse completely into radiative wave components or evolve into one or multiple solitons, which may be accompanied by soliton-like fission and supercontinuum generation \cite{dudley2006supercontinuum}. While these phenomena are consistent with the long-range observations in \cite{Su82}, their detailed analysis is beyond the scope of the present paper and is therefore omitted for brevity.


The observed extreme focusing dynamics are captured more accurately by the MNLSE, as previously discussed, for instance, in \cite{shemer2013peregrine,Chabchoub13}, whereas the other two observed regimes, characterized by either dispersion dominance or a balance between dispersion and nonlinearity, remain in excellent quantitative agreement with standard NLSE predictions.

Figure~\ref{fig:ampvsx} summarizes the envelope amplitude evolution as measured (dots) and computed from NLSE, cases $N\in\{0.36,0.6,0.8,1,1.2,1.4\}$ (a) - (c), or MNLSE for the strong focusing cases $N\in\{1.6,1.8\}$ (d), and underlines the accuracy of weakly nonlinear NLSE-type envelope evolution models to predict three distinct wave regimes, initiated by Gaussian wave packet profiles.

Beyond hydrodynamics, the present findings may motivate additional experiments involving the interaction of Gaussian wave envelopes with other coherent structures in dispersive nonlinear media \cite{chabchoub2015nonlinear,chabchoub2026extreme}. Furthermore, the results provide practical guidance for the appropriate selection of Gaussian wave-packet parameters in experimental studies of wave-structure interactions and controlled extreme wave generation.




\section*{Methods}
\label{sec:methods}

\subsection*{Nonlinear Wave Evolution Equations and Boundary Conditions}\phantomsection
\label{sec:theory}
The nonlinear evolution of the deep-water wave envelope $A(x,t)$ satisfying a narrowband spectrum can be described by the NLSE \citep{zakharov1968stability}. In a reference frame moving with the group velocity $c_g=\partial_k\omega$, the time-like NLSE, which propagates the wave packets along the spatial coordinate $x$ is given by \cite{osborne2010nonlinear}:
\begin{equation} \label{eq:nls}
    \p_xA + \rmi\frac{\beta_2}{2}\p_{tt}A + \rmi \gamma |A|^2A=0, 
\end{equation}
which can be derived from the Euler equations of fluid motion at the third-order of approximation in wave steepness $\varepsilon=ak$, that is, at $\mathcal{O}(\varepsilon^3)$. In deep-water the dispersion $\beta_2$ and nonlinearity parameter $\gamma$ are simply
\begin{equation}
    \beta_2=\frac{2}{g},~\gamma=k^3,
\end{equation}
where $k$ and $\omega$ are the carrier wavenumber and frequency, respectively. Note that both $\beta_2$ and $\gamma$ have the same sign, which allows the NLSE to admit bright soliton and breather solutions \cite{Zakharov70,akhmediev1985generation}. The time-like NLSE \eqref{eq:nls} can be rewritten into a dimensionless form
\begin{equation}
    \rmi\partial_\xi B + \frac{1}{2} \partial_{\tau\tau} B + |B|^2B = 0,
\end{equation}
through the transformation
\begin{equation}
    \xi = -a^2 \gamma x, \quad \tau = a \sqrt{\frac{\gamma}{\beta_2}} t, \quad \text{and} \quad B = \frac{A}{a},
\end{equation}
where $a$ is a characteristic amplitude used to nondimensionalize the system. For this dimensionless equation, the multi-soliton solution \cite{Satsuma74,mollenauer1980experimental,Chabchoub13} can be triggered using the initial condition 
\begin{equation}
\label{sol_par}
    B(\xi=0,\tau) = N \operatorname{sech}(\tau),
\end{equation}
with $N$ being a positive integer and ignoring their cumbersome parametrization. Transforming the latter expression (\ref{sol_par}) back to the dimensional form, we recover



\begin{equation}
    A(x=0,t) = Na \sech\left(\sqrt{\frac{\gamma}{\beta_2}}at\right).
\end{equation}
The steady envelope soliton boundary condition is trivially obtained for $N=1$ while for $N>1$, the Satsuma-Yajima soliton dynamics can be triggered \cite{Satsuma74,mollenauer1980experimental,Chabchoub13}. 
We aim to study the evolution of a Gaussian wave packet by assuming the same width as the steady bright soliton while varying its amplitude by a dimensionless factor $N$, which is now real and is not restricted to an integer. Following the convention in optics \cite{Agrawal13}, we define
\begin{align}\label{eq:gauss}
    &A_G(x=0,t) = Na \exp\left(-\frac{t^2}{2\sigma^2}\right),~\mathrm{with}\\
    &\sigma=\frac{T_\mathrm{FWHM}}{1.665}, ~~T_\mathrm{FWHM}=\frac{1.763}{a}\sqrt{\frac{\beta_2}{\gamma}}.\label{eq:FWHM}
\end{align}
Here, $T_\mathrm{FWHM}$ represents the full width at half maximum of the envelope intensity, i.e., $|A_G(x=0,t)|^2$.

The physical length scale can be represented by the second-order dispersion length $L_\mathrm{D}$ \cite{blanco2016pure} and the nonlinear length $L_\mathrm{NL}$ \cite{Agrawal13}, which are defined as
\begin{equation}\label{eq:ld}
    L_\mathrm{D}=\frac{\sigma^2}{\beta_2},~~L_\mathrm{NL}=\frac{1}{\gamma N^2a^2}.
\end{equation}

The ratio is then $L_\mathrm{D}/L_\mathrm{NL}=1.12N^2$. 

The boundary conditions in hydrodynamic laboratory experiments  are defined by the carrier wave modeled by a wave envelope $A_G(x=0,t)$. As has been previously shown, it is sufficient to generate the respective surface elevation at first-order of approximation, which is expressed as 

\begin{equation}
\eta(x=0,t)=\mathrm{Re}\!\left\{A(x=0,t)\exp\left(-\rmi\omega t\right)\right\}.
\end{equation}

The envelope of the measured waves can be calculated using the Hilbert-transform \cite{osborne2010nonlinear,chabchoub2016tracking}. Due to the discrete spatial resolution of the wave measurements along the tank, the wave envelope's amplitude $\left|A\right|$ was interpolated before being shown in Figs. \ref{fig:broaden}-\ref{fig:split}.

The NLSE is valid for a wave packet with a small steepness and narrow spectral bandwidth. In our experiments, particularly in the strong focusing regime, the wave packet becomes significantly steep, violating the validity constraint of the NLSE. As a consequence, the standard NLSE fails to accurately predict the wave evolution and the MNLSE \citep{Goullet11,chabchoub2016tracking} is adopted in addition to the NLSE simulations shown in Figs. \ref{fig:split} and \ref{fig:ampvsx}(d)
\begin{equation}\label{eq:mnlse}
\p_xA
+ \rmi\frac{1}{g}\p_{tt}A
+\rmi k^3 |A|^2 A
- \frac{k^3}{\omega}
\left(
6 |A|^2 \p_tA
+ 2 A \p_t|A|^2
- 2 \rmi A \mathcal{H}\left[\p_t|A|^2\right]
\right)
= 0,
\end{equation}
which includes terms up to $\mathcal{O}(\varepsilon^4)$ and provides a more accurate approximation, particularly for strongly focusing wave fields, thereby enabling more reliable validation against experiments. Physically, the higher-order terms successively account for self-steepening, nonlinear gradient corrections, and the wave-induced mean flow, approximated here by the Hilbert transform term $\mathcal{H}$ \cite{gomel2023mean}. The MNLSE introduces the characteristic asymmetry and accelerates the evolution of the wave envelope compared to the NLSE.
In this study, we employ the fourth-order Runge-Kutta method to integrate the NLSE and the MNLSE forward in space. The temporal derivatives were computed using a pseudo-spectral method.

\subsection*{Experimental Setup and Data Processing}
The experiments have been conducted in a large water tank facility with the dimensions 50 m $\times$ 10 m $\times$ 5 m; a schematic of the wave facility can be found in \cite{chabchoub2019directional}. The segmented wave maker can generate unidirectional and directional waves. Since our main interest is the investigation of unidirectional wave packets, all plunging wave makers generate the same wave profile, as determined by the boundary condition along the transverse tank direction. Capacitance wave gauges with a sampling frequency of 100 Hz have been placed along the tank and rearranged at different locations to improve the spatial resolution of wave measurement. We first apply a bandpass filter to remove higher Stokes harmonics followed by the Hilbert transform to obtain the wave envelope profiles \cite{osborne2010nonlinear}. The latter are then interpolated to allow for a one-to-one comparison with the numerical simulations. 
\marginpar{\footnotesize }

\section*{Data availability}
All data supporting the findings of this study are available from the corresponding
author upon reasonable request.
\section*{Code availability}
The codes for the analysis of this study are available from the corresponding author
upon reasonable request.

\bibliography{sample}
\section*{Acknowledgements}
The experiments were conducted in the Ocean Engineering Basin at the Institute of Industrial Science, the University of Tokyo. This research was supported by the Australian Research Council (ARC) Center of Excellence in Optical Microcombs for Breakthrough
Science (project no. CE230100006), funded by the Australian Government. C.M.deS is also supported by an ARC Discovery Project
(DP230102200). T.W. is supported by JSPS KAKENHI grants and JP22H00241 and JP25K24700. A.C. acknowledges support from Okinawa Institute of Science and Technology (OIST) with subsidy funding from the Cabinet Office, Government of Japan. 
\section*{Author contributions}
C.M.deS., A.B.-R., and A.C. conceived the study. A.C. and T.W. performed the experiments. Z.Z. carried out the numerical simulations and analysed the data. All authors interpreted the results and contributed to the writing and revision of the manuscript.
\end{document}